\documentstyle{article}
\def\be{\begin{equation}}
\def\ee{\end{equation}}
\def\bea{\begin{eqnarray}}
\def\eea{\end{eqnarray}}
\def\beaN{\begin{eqnarray*}}
\def\eeaN{\end{eqnarray*}}
\def\ed{\end{document}}
\def\bit{\begin{itemize}}
\def\eit{\end{itemize}}

\def\del{\delta}

\def\k{\kappa}
\def\alf{\alpha}

\def\di{\partial}

\def\half{{\textstyle{1 \over 2}}}
\def\~{\tilde}

\def\m{\label}
\def\l{\left}
\def\r{\right}

\def\goto{\rightarrow}
\def\Bar{\overline}

\begin{document}

{
\centerline{
 \bf THE SCHWARZSCHILD BLACK HOLE AS A POINT
PARTICLE}
\medskip

\centerline{\it A.~N.~Petrov}
\medskip

\centerline{Relativistic Astrophysics group, Sternberg Astronomical
institute,} \centerline{Universitetskii pr. 13, Moscow 119992,
RUSSIA}

 \centerline{ e-mail: anpetrov@rol.ru}


\begin{abstract}

The description of a point mass in general relativity (GR) is given
in the framework of the field formulation of GR where all the
dynamical fields, including the gravitational field, are considered
in a fixed background spacetime. With the use of stationary (not
static) coordinates non-singular at the horizon, the Schwarzschild
solution is presented as a point-like field configuration in a whole
background Minkowski space. The requirement of a stable
$\eta$-causality stated recently in [J.~B.~Pitts and W.~C.~Schieve,
{\em Found. Phys.} {\bf 34}, 211 (2004)] is used essentially as a
criterion  for testing configurations.

\end{abstract}

\medskip

\bigskip

\noindent{\bf 1. INTRODUCTION AND MOTIVATION}

\bigskip
 During many decades up to the present \cite{Hawking},  in numerous
classical and quantum applications and developments, the
Schwarzschild solution is the one of the most popular models in
general relativity (GR). Usually the Schwarzschild solution is
treated as a point mass solution in GR \cite{LL}. However, if one
considers GR in the usual geometrical description, then this
interpretation meets conceptual difficulties (for details see the
paper by Narlikar \cite{Narlikar} and a discussion in the paper
\cite{PN}).

Such difficulties do not appear in Newtonian gravity, where a
description of the distribution of masses and energy is very simple.
The unique Poisson equation for the gravitational potential is
considered within the matter and outside the matter. One can use
integration over both the surface surrounding  a source and  the
whole physical volume. The same formulae can be applied  both to a
continuous distribution and to a {\it point} mass. To describe a
point particle one has to assume that a distribution has the form
$m\delta({\bf r})$ where  $\delta$-function satisfies the ordinary
Poisson equation, which in spherical coordinates is
 \be
 \nabla^2 \left({1\over r}\right) \equiv \l({{d^2}\over {dr^2}} +
 {2\over r}\,{d \over {dr}}\r){1\over r} = - 4\pi\delta({\bf r})\, .
 \m{Poisson}
 \ee
Then, the Newtonian potential will apply to the whole space
including the point $r = 0$.

In \cite{PN}, it was shown that, analogously to the Newtonian
prescription, the point mass in GR can be described in a
non-contradictory manner in the framework of a so-called field
theoretical formulation (or simply ``field formulation'') of GR,
where all the dynamical fields, including the gravitational field,
are considered in a  background (fixed, auxiliary) spacetime (curved
or flat). The field formulation was developed in
\cite{GPP}~-~\cite{Petrov93} and is based on the famous paper by
Deser \cite{Deser}, who has generalized the results of previous
works of other authors deriving GR from the postulates of special
relativity (see, for example, the paper by Kraichnan
\cite{Kraichnan} as one of important papers). The field formulation
is four-covariant and is very similar to a gauge invariant field
theory in a fixed spacetime. At the same time, the field description
can be constructed with the help of a {\em simple} decomposition of
the variables of the geometrical formulation into a sum of
background and dynamical variables of the field formulation
\cite{PP}. Therefore, any solutions to GR can be treated in the
framework of the field formulation, both the formulations of GR are
equivalent locally, and they  have to be equivalent in all the
physical predictions. On the other hand, in the general case, a
manifold which supports a physical metric has not to coincide with a
manifold which supports a background auxiliary metric. As a result,
non-physical ``singularities'', ``membranes'', ``absolute voids'',
{\em etc}., can appear in a field configuration propagating on the
background. This can lead to cumbersome explanations, confused
interpretations, {\em etc.} Considering the field formulation as a
convenient tool for a resolution of several theoretical problems in
GR it is reasonable to avoid such difficulties. Thus, here we
exploit the model when a spacetime of the standard Schwarzschild
solution and a background Minkowski space are in one-to-one
correspondence.

The existence of the energy-momentum tensor (not pseudotensor) for
the gravitational field and its matter sources is one of the
advantages of the field formulation. This is the main reason why
this formulation was used in \cite{PN} to consider the energy
problem in GR. In particular, in \cite{PN} the Schwarzschild
solution was presented as a gravitational field configuration in a
background Minkowski space presented and described by the spherical
Schwarzschild ({\it static}) coordinates. The concept of Minkowski
space was extended from spatial infinity (frame of reference of a
distant observer) up to the horizon $r = r_g$ (in the Schwarzschild
coordinate r), and even under the horizon including the worldline $r
= 0$ of the true singularity. Then, the energy-momentum tensor was
constructed, the energy distribution and the total energy with
respect to the background were obtained. The configuration satisfies
the Einstein equations at all the points of the Minkowski space,
including $r = 0$. The energy distribution is presented by an
expression proportional to $\del({\bf r})$ and by free gravitational
field outside $r = 0$. The picture is clearly interpreted as a point
particle distribution in GR. Indeed, the configuration is
essentially presented by $\delta$-function, one can use the volume
integration over the {\it whole} Minkowski space and obtain the
total energy $mc^2$ in the natural way. In spite of advantages, the
interpretation of the point mass in \cite{PN} has open questions. At
$r= r_g$ both the gravitational potentials and the energy density
have discontinuities. This highlights the fact that in the standard
formulation of GR one has a coordinate singularity at $r = r_g$ in
the Schwarzschild coordinates. It is not a real singularity, and in
the field formulation  this break is interpreted as a ``bad'' fixing
of  gauge freedom. Nevetherless, a ``visible'' boundary between the
regions outside and inside the horizon exists and does not allow to
consider an evolution of events continuously.

Thus, the gauge fixing has to be improved. That is the break at $r=
r_g$ has to be countered with the use of an appropriate choice of a
flat background, which is determined by related coordinates for the
Schwarzschild solution. At least, the use of the coordinates without
singularities at the horizon, like Novikov's, Kruskal-Szekeres's,
{\it etc.}, coordinates \cite{LL,MTW}, could resolve the problem
locally at neighborhood of $r= r_g$. Besides, we restrict ourself by
the following. First, we represent a point particle at rest and in
the whole Minkowski space; therefore it has to be natural  to
describe the true singularity by the world line $r=0$ of the chosen
polar coordinates. Second, the Schwarzschild solution in appropriate
coordinates has to be asymptotically flat.

Third, we require a fulfilment of a so-called ``$\eta$-causality''
(property, when the physical light cone is inside the flat light
cone) at all the points of the Minkowski space. It is necessary to
avoid interpretation difficulties under the field theoretical
presentation of GR. By this requirement all the causally connected
events in the physical spacetime are described by the right causal
structure of the Minkowski space. A related position of the light
cones is not gauge invariant. Properties of the $\eta$-causality and
gauge transformations conserving it were studied in detail recently
by Pitts and Schieve \cite{PittsSchieve}. We take the third
requirement {\em only} to construct a more convenient in
applications and interpretation field configuration for the
Schwarzschild solution. To avoid ambiguities we stress again that,
unlike Pitts and Schieve who gives a real sense to the background,
we use it as an {\it auxiliary} construction. Thus, we agree with
the assertion by Grishchuk \cite{Grishchuk} that changing the mutual
disposition of the light cones one cannot change the physical
properties of the solution. The requirement of the $\eta$-causality
can be strengthened   by the requirement of a ``stable
$\eta$-causality'' \cite{PittsSchieve}. The last means that the
physical light cone has to be {\it strictly} inside the flat light
cone, and  this is important when quantization problems are under
consideration. Indeed, in the case of tangency a field is on the
verge of $\eta$-causality violation \cite{PittsSchieve}. Returning
to the presentation in the Schwarzschild coordinates in \cite{PN} we
note that it does not satisfy the third requirement.

More appropriate coordinates are, first, the {\it stationary} (not
static) coordinates presented in \cite{Petrov90,Petrov92}
(independently in \cite{Vlasov,Loskutov}), and recently improved in
\cite{PittsSchieve}, second, contracting Eddingtom-Finkenstein
coordinates in stationary form \cite{MTW}. These coordinate systems
belong to a parameterized family where all of systems satisfies all
the above requirements. Qualitatively the aforementioned two systems
present all the important properties of the family (see discussion
at the end of the paper). Therefore, for the sake of simplicity and
clarity we use just them to approach the goal of the present letter,
that is to describe the Schwarzschild solution as a point particle
in GR.

Except a pure theoretical interest the description given in the
present paper could be interesting and useful for experimental
gravity problems.  Gravitational wave detectors such as LIGO and
VIRGO will definitely discover gravitational waves from coalescing
binary systems comprising of compact relativistic objects. Therefore
it is necessary to derive equations of motion of such components,
e.g., two black holes. As a rule, at an {\em initial} step the black
holes are modeled by point-like particles presented by Dirac's
$\delta$-function. Then consequent post-Newtonian approximations are
used (see the works with excellent mathematics
\cite{Damour-1}~-~\cite{Damour-4} and references therein). However
this approach meets difficulties related to the non-linear nature of
the Einstein equations. Different regularization methods have been
suggested to bypass them. However, in spite of a significant
progress, so far the problem of motion of the black holes in GR has
many of open questions \cite{Damour-1}~-~\cite{Damour-4}. Our way of
definition of a point-like source of gravity  is different. Not
making {\em initial} assumptions on its structure we use the
Schwarzschild solution itself to define it. A resulting field
configuration including a description of the true singularity in the
the form of a point-like particle is easy for applications and
allows to reproduce the Schwarzschild solution as is --- without
approximations, with correctly-defined position of the horizon, {\em
etc.}

\bigskip
\bigskip
\noindent{\bf 2. ELEMENTS OF THE FIELD FORMULATION OF GR}
\bigskip

 At first, we briefly repeat the main notions of the field
formulation of GR \cite{GPP}. Here, it is enough to consider the
equations for the gravitational field $h^{\mu\nu}$ on Ricci-flat
backgrounds:
 \be
G^L_{\mu\nu}(h^{\alpha\beta}) = {\kappa {t^{tot}_{\mu\nu}}}\, .
 \m{FieldEqs}
 \ee
The left hand side is linear in the symmetric tensor $h^{\mu\nu}$:
 \be
G^L_{\mu\nu}(h^{\alpha\beta}) \equiv {\half
\l(h^{~~~;\alpha}_{\mu\nu~~;\alpha} +
\gamma_{\mu\nu}h^{\alpha\beta}_{~~~;\alpha\beta} -
h^\alpha_{~\mu;\nu\alpha} - h^\alpha_{~\nu;\mu\alpha}\r)}\,
 \m{GL}
 \ee
where  $\gamma_{\mu\nu}$ is the background metric; $\gamma \equiv
{\det {\gamma_{\mu\nu}}}$; $(;\alpha)$ means the covariant
derivative with respect to $\gamma_{\mu\nu}$. The total
energy-momentum tensor
 \be
t^{tot}_{\mu\nu} \equiv {t^g_{\mu\nu} + t^m_{\mu\nu}}
 \m{t-tot}
 \ee
is  obtained after varying the action of GR in the field form with
respect to $\gamma^{\mu\nu}$. The pure gravitational part of
(\ref{t-tot}) has the form:
 \be
\kappa t^g_{\mu\nu} = -(KK)_{\mu\nu} + \half \gamma_{\mu\nu}
(KK)^{~\alpha}_\alpha + Q^\sigma_{~\mu\nu;\sigma}
 \m{t-g}
 \ee
 with the tensors
 \bea
 (KK)_{\mu\nu}& \equiv &{K^\alpha_{~\mu\nu}K^\beta_{~\beta\alpha} -
 K^\alpha_{~\mu\beta}K^\beta_{~\nu\alpha}}\, ,
 \m{KK}\\
Q^{\sigma}_{~\mu\nu}& \equiv & {-
\half\gamma_{\mu\nu}h^{\alpha\beta} K^\sigma_{~\alpha\beta}} + \half
{h_{\mu\nu}K^{\alpha~\sigma}_{~\alpha}}
 - {h^{\sigma}{}_{(\mu}K^{\alpha}_{~\nu)\alpha}}\nonumber\\ & +&
h^{\beta\sigma}K^\alpha_{~\beta(\mu}\gamma_{\nu)\alpha} +
h^{\beta}{}_{(\mu}K^\sigma_{~\nu)\beta} - h^{\beta}{}_{(\mu}
\gamma_{\nu)\alpha}K^\alpha_{~\beta\rho}\gamma^{\rho\sigma}\, ,
 \m{Q}\\
K^\alpha_{~\beta\gamma} &\equiv & \Gamma^\alpha_{~\beta\gamma} -
 C^\alpha_{~\beta\gamma}\,
 \m{K}
 \eea
where  $\Gamma^\alpha_{~\beta\gamma}$ and $C^\alpha_{~\beta\gamma}$
are the Christoffel symbols for the dynamic (physical) and
background spacetimes respectively. Note that in fact the field
configuration is defined by the components $h^{\mu\nu}$. However,
sometimes variables of the 1-st order formalism are more convenient,
thus in expressions (\ref{t-g}) - (\ref{Q}) the components of the
tensors $h^{\mu\nu}$ and $K^\alpha_{~\beta\gamma} $ are used as
independent variables (see for the details \cite{PP}). Note also
that if Eq. (\ref{FieldEqs}) is satisfied, then the total
energy-momentum tensor (\ref{t-tot}) can be obtained with the use of
its left hand side, that is with the expression (\ref{GL}).

The equivalence between the field and the geometrical formulations
 of GR can be stated after the
 simple identifications
 \bea
  \sqrt{-\gamma}\l(\gamma^{\mu\nu} + h^{\mu\nu}\r) &\equiv &
  \sqrt{-g}g^{\mu\nu}\,\nonumber\\
C^\alpha_{~\beta\gamma} + K^\alpha_{~\beta\gamma} & \equiv &
\Gamma^\alpha_{~\beta\gamma} = \half
g^{\alpha\rho}\l(g_{\rho\beta,\gamma} + g_{\rho\gamma,\beta} -
g_{\beta\gamma,\rho}\r)
 \m{identification}
 \eea
where $g \equiv \det {g_{\mu\nu}}$. Then, the  equations
(\ref{FieldEqs}) change over to the usual form of the Einstein
equations with the dynamic metric ${g_{\mu\nu}}$. The source
energy-momentum tensor in (\ref{t-tot})  is connected with the usual
matter energy-momentum tensor $T_{\mu\nu}$ of GR  as
 \be
t^m_{\mu\nu} = T_{\mu\nu} - \half
g_{\mu\nu}T_{\alpha\beta}g^{\alpha\beta} - \half
\gamma_{\mu\nu}\gamma^{\alpha\beta}\l(T_{\alpha\beta} - \half
g_{\alpha\beta}T_{\pi\rho}g^{\pi\rho}\r)\, .
 \m{t-T}
 \ee

\bigskip
\bigskip

\noindent{\bf 3. THE SCHWARZSCHILD SOLUTION IN A STABLY {\LARGE $
{\bf \eta}$}-CAUSAL DESCRIPTION AND THE TRUE SINGULARITY}

\bigskip

A Minkowski space related to the stationary coordinates $\{t,\,
r^*,\,\theta,\, \phi \}$ constructed in \cite{Petrov90,Petrov92} for
the Schwarzschild solution  does not cover the region around the
true singularity with the radius less $r_g/2$. After making a
translation of the radial coordinate $r^* \goto r= r^* + r_g/2$, as
it was suggested in \cite{PittsSchieve}, a corresponding Minkowski
space just covers the whole region of the {\em standard}
Schwarzschild solution under the horizon including the singularity
at $r= 0$. Thus, the stationary metric \cite{Petrov90,Petrov92} gets
the modified form \cite{PittsSchieve}:
 \be
 ds^2 = \l(1 - \frac{r_g}{r}\r)c^2dt^2 - 2\,\frac{r_g^2}{r^2}\, c\, dt\, dr
 - \l(1 + \frac{r_g}{r}\r) \l(1 + \frac{r_g^2}{r^2}\r)dr^2 -
 r^2\l(d\theta^2 + \sin^2\theta d\phi^2\r)\,.
 \m{stationary}
 \ee
Coordinates $\{t,\, r,\,\theta,\, \phi \}$ of this metric are
connected with the standard Schwarzschild coordinates $\{T,\,
r,\,\theta,\, \phi \}$ by the transformation of the time coordinate
only:
 \be
 ct = cT + r_g\ln\l|1 - \frac{r_g}{r}\r|\, .
 \m{T-to-t}
 \ee

The important property of the solution (\ref{stationary}) is that a
falling test particle reaches the horizon $r= r_g$ in finite
coordinate time $t$, under the horizon is always falling towards the
singularity, gets arbitrarily close to it, but only hits it at $t=
\infty$ (see \cite{Petrov92}). However, in Minkowski space there are
simply no events with $t \ge \infty$, as it was noted in
\cite{PittsSchieve}.

As is seen, the metric (\ref{stationary}) is left stationary due to
the non-zero cross component $g_{01} = r_g^2/r^2$. Thus, analogously
to the Kerr solution \cite{LL,MTW} that presents the rotating
dragging, or to the Lorentz transformed Schwarzschild solution
\cite{BHMembranes} that presents the dragging in the direction of a
velocity of distant observer, the solution (\ref{stationary})
presents the dragging in the direction of the singularity. In this
respect it is a place to note the Gullstrand-Painleve form of the
Schwarzschild solution (see, e.g., a recent paper \cite{river}). It
is connected with the standard Schwarzschild metric by the
transformation $ ct_{GP} = cT + r_g\l(2\beta + \ln
\l|(1+\beta)/(1-\beta)\r|\r)$ with $\beta = (r_g/r)^{1/2}$ and is
similar to (\ref{stationary}). The Gullstrand-Painleve metric is
also stationary and presents the dragging directed to the
singularity. The last property is used to conceptualize a black hole
as a river model \cite{river}: the space itself flows like a river
through a flat background, while objects move through the river
according to the rules of special relativity. But this solution
cannot be considered here because it does not satisfy to the third
($\eta$-causality) requirement.

Now let us present the solution (\ref{stationary}) in the form of a
field configuration in the Minkowski space with the metric in  the
polar coordinates:
 \be
 d\Bar s^2 =c^2dt^2 - dr^2 -
 r^2\l(d\theta^2 + \sin^2\theta d\phi^2\r)\,
 \m{fon}
 \ee
where we will numerate the coordinates as $x^0 = ct$, $x^1 = r$,
$x^2 = \theta$ and $x^3 = \phi$. The use of the physical metric
(\ref{stationary}) and the background metric (\ref{fon}) in the
relations  (\ref{identification}) and (\ref{K}) give a possibility
to construct the field configuration
 \be
 h^{00}  =  \frac{r_g}{r}+ \frac{r_g^2}{r^2} + \frac{r_g^3}{r^3}\,
 ,\qquad h^{01}  =  - \frac{r_g^2}{r^2}\, ,\qquad
 h^{11}  =  \frac{r_g}{r}\, ;
 \m{h-configur}
 \ee
\bea
 K^0_{~00} & = & - K^1_{~01} = \frac{1}{2}\,\frac{r_g^3}{r^4}\, ,
\nonumber\\
 K^0_{~01} & = & - K^1_{~11} = \frac{1}{2}\,\frac{r_g}{r^3}
 \l(1 + \frac{r_g}{r}\r) \l(1 + \frac{r_g^2}{r^2}\r)\, ,
\nonumber\\
 K^0_{~11} & = & \frac{1}{2}\,\frac{r_g^2}{r^3}
 \l(4 + 3\frac{r_g}{r}+ 2\frac{r_g^2}{r^2} + \frac{r_g^3}{r^3}\r)\, ,
\nonumber\\
K^1_{~00} & = &   \frac{1}{2}\,\frac{r_g}{r^2}
 \l(1 - \frac{r_g}{r}\r) \, ,
 \nonumber\\
K^0_{~33} & = & K^0_{~22} \sin^2 \theta  = - \frac{r_g^2}{r}\sin^2
\theta\, ,
\nonumber\\
K^1_{~33} & = & K^1_{~22} \sin^2 \theta  = r_g\sin^2 \theta\, .
 \m{K-configur}
 \eea

Now we assume that the field configuration (\ref{h-configur}) and
(\ref{K-configur}) satisfies the Einstein equations (\ref{FieldEqs})
at every point of Minkowski space, including $r= 0$. Then for
calculating the components of the energy-momentum tensor
(\ref{t-tot}) and its parts it is important to define the expression
$\nabla^2(1/r^{k+1})$ with integer $k \ge 0$. Recalling that for
$k=0$ it is already given in (\ref{Poisson}), we use the technique
of the generalized functions \cite{GSh}. Thus, considering the
expression
 \be
 \Phi_i = \frac{\di}{\di x^i}\frac{1}{r^{k+1}} = -(k+1)\frac{x_i}{r^{k+3}}
 \label{Phii}
 \ee
as a generalized  homogeneous function of the $-k-2$ degree in $3$
dimensions one can apply to it the rules of a differentiation
derived in \cite{Kopeikin} and based on the standard notions
\cite{GSh}. This gives
 \be
\nabla^2\frac{1}{r^{k+1}}= (k+1)\l[ \frac{k}{r^{k+3}}
-\frac{(-1)^{k}}{k!} \frac{\di^{k} \delta({\bf r})} {\di r^{k}}
{n_{\alf_1}\ldots n_{\alf_k}} \oint_\Gamma
 n^{\alf_1}\ldots n^{\alf_k}d \Omega\r]\, .
 \label{Delta-rk+}
 \ee
In this paragraph we use the related Cartesian coordinates
$\{x^i\}$, $i=1,\,2,\,3$, with that $n^i = x^i/r$; $\Gamma$ is a
closed two-surface surrounding a singular point; $d\Omega = r^{-2}
n^i ds_i$ where $ds_i$ is the element of integration on $\Gamma$. On
the other hand, one can consider $\Psi_i = r^{k} \Phi_i$ as a
generalized  homogeneous function of the $-2$ degree and apply the
rule of differentiation  given in \cite{GSh} to $\Psi_i$. The final
expression  is
 \be
 \nabla^2\frac{1}{r^{k+1}}= (k+1)\l[ \frac{k}{r^{k+3}}
-\frac{4\pi}{r^{k}}  \delta({\bf r})\r]\, .
 \label{Delta-rk++}
 \ee
Comparing (\ref{Delta-rk+}) with (\ref{Delta-rk++}) one finds an
equivalence between the last terms in these formulae. Reducing this
equivalence to the simplest case of $1$ dimension, for example, one
obtains the known relation \cite{Korn}: $\di^{k}\delta(x)/\di x^k =
(-1)^k k!\,x^{-k} \delta(x)$. Here we prefer to use the formula
(\ref{Delta-rk++}) as a more convenient in calculations. Thus, e.g.,
it is easy to see that integration over a round ball of the r.h.s.
of (\ref{Delta-rk++}) gives two divergent integrals at $r \goto 0$
that compensate one another. Then, a convergent part of this volume
integral is equal to a value of a surface integral that follows
after integration of the l.h.s. of (\ref{Delta-rk++}), which is a
divergence $\nabla^2 = \di_i\di^i$. In \cite{PN} we use also the
presentation (\ref{Delta-rk++}).

For the calculations of $t^{tot}_{\mu\nu}$ we use the expression
(\ref{GL}), the non-zero components of which are
 \bea
 t^{tot}_{00} & = & mc^2 \delta({\bf r})+  mc^2\,\frac{r_g}{r}\l(1 +
 \frac{3}{2}\frac{r_g}{r}\r)\delta({\bf r}) - \frac{mc^2}{4\pi}\,\frac{r_g}{r^4}
\l(1+ 3\frac{r_g}{r}\r)\, ,\nonumber\\
t^{tot}_{11} & = & - mc^2\delta({\bf r})\, ,\nonumber\\
t^{tot}_{AB} & = & - \half \gamma_{AB}\, mc^2\delta({\bf
r})\,;\qquad A,\,B = 2, \, 3.
 \m{t-tot-q}
 \eea
For calculations of the free gravitational part in (\ref{t-tot}) we
use the expressions (\ref{t-g}) - (\ref{Q}):
 \bea
 t^{g}_{00} & = & {mc^2}\,\frac{r_g}{4r}\, \l(6+ 7\frac{r_g}{r}+
 \frac{r_g^2}{r^2}\r)\delta({\bf r}) - \frac{mc^2}{4\pi}\,\frac{r_g}{r^4}
\l(1+ 3\frac{r_g}{r}\r)\, ,\nonumber\\
t^{g}_{01} & = & {mc^2}\,\frac{r_g^2}{2r^2}\,\,\delta({\bf r})\, ,\nonumber\\
t^{g}_{11} & = & {mc^2}\,\frac{r_g}{2r}\, \l(1+ \frac{r_g}{2r}+
 \frac{r_g^2}{2r^2}\r)\delta({\bf r})\, ,\nonumber\\
t^{g}_{AB} & = & \gamma_{AB}\,{mc^2}\,\frac{r_g^2}{4r^2}\, \l(1+
\frac{r_g}{r}\r)\delta({\bf r})\,.
 \m{t-g-q}
 \eea
In calculations of the components (\ref{t-tot-q}) and (\ref{t-g-q})
it was used the usual notations $\kappa = 8\pi G/c^4$ and $r_g =
2mG/c^2$. Now, for the calculation of $t^{m}_{\mu\nu}$ of the matter
part we use the difference between (\ref{t-tot-q}) and
(\ref{t-g-q}):
 \bea
 t^{m}_{00} & = & mc^2 \delta({\bf r}) - {mc^2}\,\frac{r_g}{2r}\, \l(1+ \frac{r_g}{2r}+
 \frac{r_g^2}{2r^2}\r)\delta({\bf r}) \, ,\nonumber\\
t^{m}_{01} & = & - {mc^2}\,\frac{r_g^2}{2r^2}\,\,\delta({\bf r})\, ,\nonumber\\
t^{m}_{11} & = & - mc^2 \delta({\bf r}) - {mc^2}\,\frac{r_g}{2r}\,
\l(1+ \frac{r_g}{2r}+
 \frac{r_g^2}{2r^2}\r)\delta({\bf r}) \, ,\nonumber\\
t^{m}_{AB} & = & - \half \gamma_{AB}\,{mc^2}\l(1+
\frac{r_g^2}{2r^2}+ \frac{r_g^3}{2r^3}\r) \delta({\bf r})\,.
 \m{t-m-q}
 \eea

The components (\ref{t-m-q}) can be obtained directly. With the
physical metric (\ref{stationary}) one has to calculate the Einstein
tensor $G_{\mu\nu}$ everywhere including $r=0$ and, thus, define the
components of the matter tensor  $T_{\mu\nu}$, which could be a
source for $G_{\mu\nu}$. Then with  using the relation (\ref{t-T})
the components (\ref{t-m-q}) are obtained again. However one has to
note, in this case components $T_{\mu\nu}$ obtained in the framework
of the ordinary geometrical formulation of GR do not have a good
interpretation \cite{Narlikar}.

Let us discuss properties of the field presentation of the solution
(\ref{stationary}). First, it is in the spirit of GR that
$t^m_{\mu\nu}$ can not be considered separately from $t^g_{\mu\nu}$.
Thus, it is more right to consider the total components
(\ref{t-tot-q}). The energy distribution is described by the
$00$-component of the energy-momentum tensor. Then the total energy
of the system is obtained with the use of the volume integration:
 \be
 E^{tot} = \lim_{r \goto \infty} \int_V t^{tot}_{00}r^2\sin\theta dr
 d \theta d \phi = mc^2\, .
 \m{int-V}
 \ee
It is defined only by the first term $mc^2\, \delta({\bf r})$ in
$t^{tot}_{00}$ that follows from  the matter component $t^m_{00}$
only. The other contributions into (\ref{int-V}) from the
$\delta$-functions in $t^{tot}_{00}$ are infinite, but they are
compensated by the energy distribution without $\delta$-functions
that is a part of the gravitational component $t^g_{00}$. Due to
(\ref{FieldEqs}) the volume integration can be exchanged by the
surface integration over the 2-sphere with the constant $r = r_0$:
 \be
 E^{tot} = \lim_{r_0 \goto \infty}\frac{1}{2\k} \oint_{\di V}
 \l(h_{00}{}^{;1} + \gamma_{00} h^{1\alpha}{}_{;\alpha} - 2h^{1}{}_{0;0} \r)
 r^2\sin\theta d \theta d \phi = mc^2\, .
 \m{int-diV}
 \ee
The other components $t^{tot}_{11}$ and $t^{tot}_{AB}$ in
(\ref{t-tot-q}) formally could be interpreted as related to the
``inner'' properties of the point. Indeed, they are proportional
only to $\delta({\bf r})$ and, thus, describe the point ``inner
radial'' and ``inner tangent'' pressure.

Second, after transformation from the spherical coordinates in
(\ref{stationary}) to the corresponding Cartesian coordinates one
can see that the metric (\ref{stationary}) and the configuration
(\ref{h-configur}) are asymptotically flat with the $1/r$-like
falloff at spatial infinity. As it was stated in \cite{Petrov95},
where the gauge invariance of integrals of motion of an isolated
system was studied, the $1/r$-like  asymptotic behaviour just
guarantees the satisfactory results (\ref{int-V}) and
(\ref{int-diV}) for the total energy. Third, the metrics
(\ref{stationary}) and (\ref{fon})  satisfy the requirement of the
{\it stable $\eta$-causality} at all the points of the Minkowski
space down to the true singularity at $r= 0$. Thus, all the
requirements are satisfied.

The presented picture is more complicated than in the case of the
point mass in the Newtonian gravity. Nevetherless, the problem of
the point mass is resolved enough simply. Indeed, the
energy-momentum tensors contain $\delta$-functions at $r = 0$, and,
like in the Newtonian case, the volume integration over the whole
space gives a satisfactory total energy. On the other hand, the
presented here description is significantly simpler and more
appropriate than in \cite{PN}. The field configuration
(\ref{h-configur}), unlike \cite{PN}, is continuous at {\it all} the
point of the Minkowski space except the true singularity $r=0$, that
is natural. A falling test particle approaches and intersects the
horizon $r= r_g$ in finite Mikowski time $t$. The components
$t^{tot}_{00}$ and $t^g_{00}$ have no breaks outside $r=0$, and all
the other energy-momentum components in (\ref{t-tot-q}) -
(\ref{t-m-q}) are defined only by a $\delta$-function.

\bigskip
\bigskip

\noindent{\bf 4. A FIELD THEORETICAL REFORMULATION OF THE
CONTRACTING EDDINGTON-FINKELSTEIN METRIC}

\bigskip

Now let us examine the contracting Eddington-Finkelstein metric for
the Schwarzschild geometry \cite{MTW}:
 \be
 ds^2 = \l(1 - \frac{r_g}{r}\r)c^2d\~t^2 - 2\,\frac{r_g}{r}\, c\, d\~t\, dr
 - \l(1 + \frac{r_g}{r}\r)dr^2 -
 r^2\l(d\theta^2 + \sin^2\theta d\phi^2\r)\,.
 \m{EF}
 \ee
Notice that a transformation was made from the standard null
coordinate $\~V$ to time coordinate $\~t$: $\~t = \~V - r$. If the
flat background, analogously to (\ref{fon}), is described by the
coordinates $c\~t$, $r$, $\theta$ and $\phi$, then the gravitational
field configuration corresponding to (\ref{EF}) is
 \be
 h^{00}  =  \frac{r_g}{r}\,
 ,\qquad h^{01}  =  - \frac{r_g}{r}\, ,\qquad
 h^{11}  =  \frac{r_g}{r}\, .
  \m{h-configur-EF}
 \ee
The properties of the solutions (\ref{stationary}) and (\ref{EF})
are very close. Both metrics are stationary and asymptotically flat.
In the whole Minkowski space they induce asymptotically flat and
continuous (except $r=0$) configurations (\ref{h-configur}) and
(\ref{h-configur-EF}). Falling test particles intersect the horizon
$r= r_g$ in finite times $t$ and $\~t$, but in the the case
(\ref{EF}) test particles even reach the true singularity in finite
time $\~t$. This is the result of the time transformation for
(\ref{EF}) \cite{MTW}: $c\~t = cT + r_g\ln\l|1 - {r}/{r_g}\r|$
instead of (\ref{T-to-t}).

The components of the total energy-momentum tensor for the
configuration (\ref{h-configur-EF}) are \bea
 t^{tot}_{00} & = & mc^2 \delta({\bf r})\, ,\nonumber\\
t^{tot}_{11} & = & - mc^2\delta({\bf r})\, ,\nonumber\\
t^{tot}_{AB} & = & - \half \gamma_{AB}\, mc^2\delta({\bf r})\,.
 \m{t-tot-EF}
 \eea
This energy-momentum, unlike (\ref{t-tot-q}), is concentrated {\it
only} at $r=0$. Of course, the volume integration of $t^{tot}_{00}$
from (\ref{t-tot-EF}) again, like (\ref{int-V}), gives $E^{tot} =
mc^2$, and the surface integration (\ref{int-diV}) with the
configuration (\ref{h-configur-EF}) gives it also. However, unlike
(\ref{int-diV}), now  $mc^2$ follows with arbitrary radius of the
2-sphere $r_0$ (it is not necessary $r_0 \goto \infty$), like for
the electric charge in electrodynamics and for the point mass in
Newtonian gravity. This situation is very close to the Penrose
charge integral prescription \cite{Penrose} for the ``quasi-local
mass'' $m_P = m(\di V)$ surrounded by a 2-sphere $\di V$.  Tod
\cite{Tod} has adopted the Penrose construction for 2-surfaces of
spherical symmetry in spherically symmetric spacetimes. Thus, the
Schwarzschild mass parameter $m = m_P$ is obtained {\it
independently} of radius of $\di V$. As is seen, with the solution
(\ref{EF}) the description of the point mass in GR looks also quite
appropriate.

The transformation $ct' = cT + r_g \ln |(r/r_g- 1)(r_g/r)^\alf|$
gives a parameterized by $\alf \in [0,2]$ family of metrics, all of
which satisfies all our requirements, the cases $\alf = 0$ and $\alf
= 1$ correspond to (\ref{EF}) and (\ref{stationary}). At this the
requirement of the {\em stable} $\eta$-causality {\em is not}
satisfied with $\alf = 0$ at $0 \le r \le \infty$. Thus, all the
configurations $\alf \in (0,2]$ are appropriate for the study both
classical and quantum problems, whereas the case $\alf = 0$ could
not be useful for the study quantized fields. Properties of field
configurations corresponding to  $\alf \in (0,2]$ qualitatively are
the same as for $\alf = 1$. In the terms of the field approach
\cite{GPP}, all the field configurations for $\alf \in [0,2]$ are
connected by gauge transformations and are physically equivalent.
Thus, inside this family, $\eta$-causal description with
(\ref{h-configur-EF}) can be converted into a stably $\eta$-causal
description explicitly. Note also that a technique of infinitesimal
gauge transformation developed in \cite{PittsSchieve} permits do
this conversion approximately without relation to this family.

 \section*{Acknowledgments}
Author expresses his gratitude to Brian Pitts for very useful and
interesting discussions and recommendations. He thanks also all the
referee, recommendations and remarks of whom helped to improve the
manuscript significantly.

\ed